# Designing a Routing Protocol for Ubiquitous Networks Using ECA Scheme

Chandrashekhar Pomu Chavan and Pallapa Venkataram

Protocol Engineering and Technology Unit, Department of Electrical Communication Engineering, Indian Institute of Science, Bangalore, India

*{cpchavan, pallapa}@ece.iisc.ernet.in*

*Abstract*

*We have designed a novel Event-Condition-Action (ECA) scheme based Ad hoc On-demand Distance Vector(ECA-AODV) routing protocol for a Ubiquitous Network (UbiNet). ECA-AODV is designed to make routing decision dynamically and quicker response to dynamic network conditions as and when event occur. ECA scheme essentially consists of three modules to make runtime routing decision quicker. First, event module receive event that occur in a UbiNet and split up event into event type and event attributes. Second, condition module obtain event details from event module split up each condition into condition attributes that matches event and fire the rule as soon as condition hold. Third, action module make runtime decisions based on event obtained and condition applied. We have simulated and tested the designed ECA scheme by considering ubiquitous museum environment as a case study with nodes range from 10 to 100. The simulation results show the time efficient with minimal operations.*

## 1. Introduction

Ubiquitous Network (UbiNet) is a heterogeneous network [1] with various computing devices which are connected at anywhere, anytime and enable users to access and exchange information [2]. In UbiNet, the nodes may join/leave the network frequently and move freely hence, leads to frequent change in network topology. Since the nodes are mobile they can move arbitrarily in any direction leads to various failures like link failure, node failure and so on. Hence the dynamic network topology and frequent failures can be addressed using a routing protocol to manage the variety of failures and to choose optimal paths to transmit the data [3].

Routing in UbiNet is to find a best path from user to the service provider. Basically, routing algorithms are designed to determine the best paths in the network, whereas routing table entries store route information that the algorithm has already discovered a best path and routing protocols allow data packet to be collected and distributed across the network [4]. A UbiNet do not support any existing routing protocols since a client node does not have(or any node) or do not maintain the routing tables[5], [6].

We develop an ECA scheme based routing protocol in UbiNet i.e. ECA-AODV routing protocol for making dynamic ubiquitous routing decision quickly at runtime. In an ECA scheme, event [7] is defined as significant changes in state of a system. Event can occur at any time, event module collect the detail information about the event and forward to condition module for applying logical procedure in order to make dynamic routing decision quickly. When a specific event is occurred at a particular time, certain conditions are met then action module make dynamic decision [8],[9],[10].

### 1.1. Proposed idea

We have proposed an ECA scheme for routing protocol in UbiNet. An ECA scheme is broadly divided into three modules namely Event module, Condition module and Action module respectively. Event module is a 2-tuples consists of event types and event attributes, the function of event module is to keep observe and notifies events. Condition module is 2-tuples consists of event details and condition attributes.

Condition module observes and receives an event, look for the rule that matches inputs and fire the rule as soon as condition hold. Based on event and condition, a decision is made by the action module which contains 1-tuple action attributes. ECA scheme is store using structure data structure and distributed across every node. Upon receiving event at a particular node, node intern broadcast occurred event to all its vicinity nodes

### 1.2. Pattern of the research paper

Pattern of the research paper is as follows. Section 2 present most relevant works. Section 3 briefly explains AODV routing. Section 4 briefly describes routing in UbiNet. Section 5 explain proposed ECA scheme. Section 6 describes the ubiquitous museum environment case study. Section 7 gives the simulation environment. 8 shows the simulation result and finally, the paper draws some conclusions in Section 9.

## 2. Some of the existing works

Bhandari S.R. and Bergmann N.W [11] describes program do not respond well for component or resource failure. An ECA based system is suitable for both describing desired system operations as well as linking an event [12] based system to communicate with resources.

Hannes Obweger et al.[13] propose an innovative framework for creating sense and response rules that can be useful for real-time. Sense and response has set of rules to detect event [14] from business stream and take appropriate decision based on event occurrence.

Kaan Bur and Cem Ersoy [15] presents a novel mesh based QoS multicast routing which keep track of the resources availability at every node and monitor the QoS status periodically. QoS multicast routing will be selected on the basis of available resources, protocol selects optimal route in heterogeneous network. Gateway node does protocol conversion at the boundaries of the subnet.

Jungil Heo and Wooshik Kim [16] analyze the information about user movement among heterogeneous subnet technologies in a ubiquitous system and proposes how to configure and manage the network in order to execute ubiquitous service in time. AODV protocol plays a vital role in health care environment to provide reliable and power efficiency data transmission.

Elizabeth M. Royer and Charles E. Perkins [17] consider fundamental functioning of AODV routing protocol in which, nodes store the route as and when needed. Each node uses destination sequence numbers to ensure freshness of the route at all times, AODV response quickly during link breakage in active routes.

## 3. AODV Routing

The AODV routing protocol is on-demand and reactive, in which path is discovered on-demand; established path is maintained as long as it is needed. AODV routing protocol [18] consists of four messages viz. i) Route request message, ii) Route reply message, iii) Route error message and iv) Route reply-acknowledgement message respectively. Essentially, node relies on intermediate nodes to find an optimal path from originating node to the ultimate target node in the network.

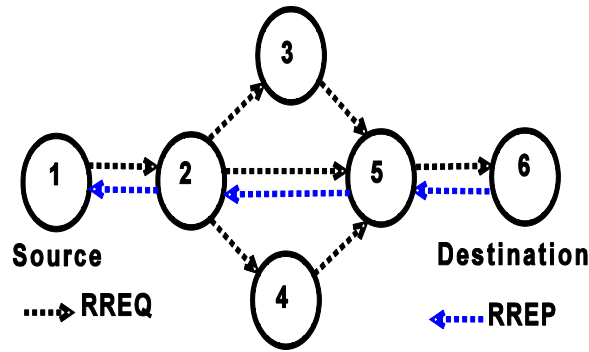

Figure 1. Route Discovery process in AODV Routing

In Figure 1, Node 1 has some data, it wishes to send to node 6 but node 1 does not have a valid path to communicate with node 6, in such case node 1 initiate path finding by sending route request message to its vicinity node 2, node 2 intern propagate the route request to its neighbour nodes such as 3, 4 and 5 respectively until route request reaches destination, if anyone of the intermediate node has the valid path toward the destination or node itself is destination may reply to the corresponding route request. Immediately upon receiving route request node verify that the sequence number [19] of the replying node is greater than that contained in route request message, then only node initiate to generate the route reply and unicast back to the route request originator. This is how a valid route is established in AODV routing.

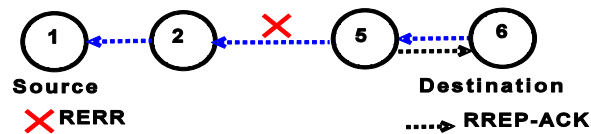

Figure 2. Route error and Route reply-ACK message

AODV uses route request and route reply for establishing a valid path, upon establishing valid path node store a routing table to communicate with rest of the nodes in the network, if existing path is not valid then node generate and forward the route error message to its predecessor node, when a node activate a bidirectional link then it send route reply acknowledgement message.

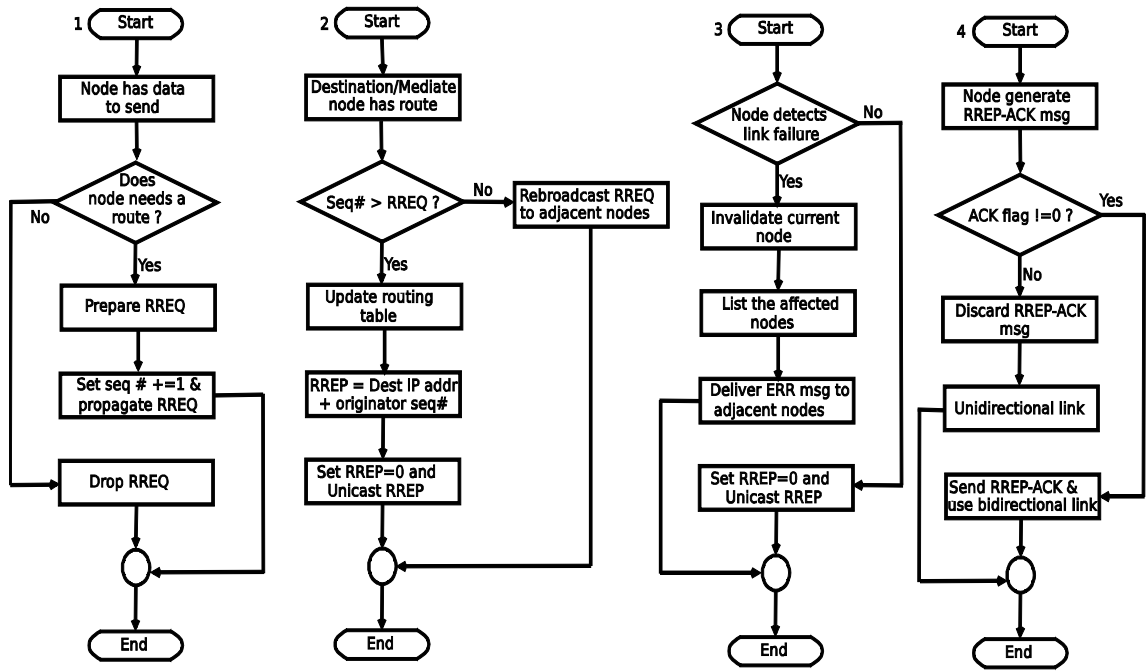

Figure 3. Flow chart of generating four AODV messages

Usually, in AODV originating node initiate a path discovery process as and when it would like to communicate with other node, node itself verify it has valid route or not if valid route does not exist then source node prepare RREQ packet and increment sequence number by one and propagate to its adjacent nodes

## 4. Routing in Ubiquitous Networks

In ubiquitous environment, users demand constant availability of service at anytime from anywhere. Ubiquitous Server (UbiServ) receive informations about the user such as location, time, types of device, interest, preference, etc. from various embedded sensors and user profile, process the information and play a vital role in providing routing information to the user. UbiServ are connected and distributed across the internet, when a user enter into ubiquitous environment, an automatic path is establishes from user to the UbiServ without user's intervention [20].

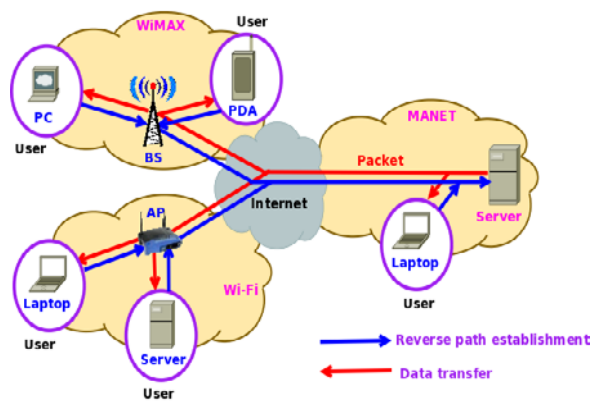

Figure 4. Example of routing in ubiquitous network

Routing in ubiquitous network has numerous differences as compared to normal routing in the following points

- ❖ Routing is adapted to heterogeneous network based on application requirements.
- ❖ UbiServ provide end-to-end flexible routing.
- ❖ Routing decision can be made dynamically based on current network status.
- ❖ Provide seamless service.
- ❖ Support QoS guaranteed service in high traffic.
- ❖ Lossless handover during switching from one network access technology to another network access technology.
- ❖ Provide best-effort traffic during congestion.

In ubiquitous network, a reverse path is established from user to the ubiquitous server, upon establishing path, UbiServ maintain an uninterrupted connectivity from user to the various subnet. UbiServ provide the routing information to the user located at heterogeneous subnet based on user's interest and preference.

## 5. Proposed ECA scheme in UbiNet

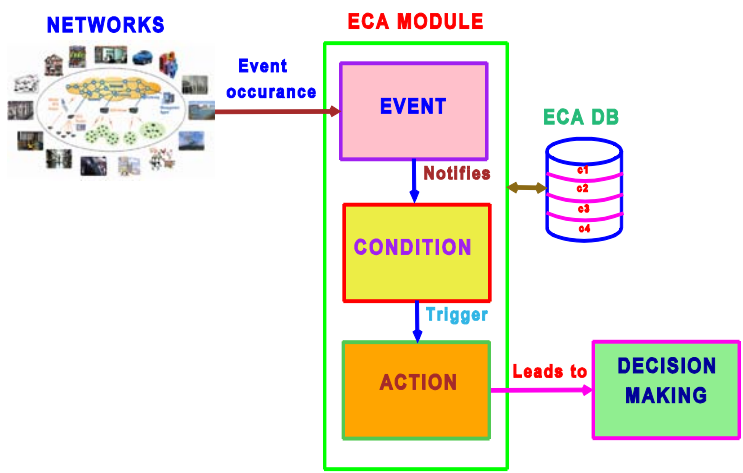

Figure 5. Function block diagram of an Event-Condition-Action Scheme

### 5.1. Function of event module

Event is defined as significant changes in state of a system. Event module keep observe events and notifies as and when events occurs. Event is a 2-tuple consists of event types and event attributes, event is denoted by $E_i$ notation, event type may be time, spatial, composite, request, notification, internal, external, fault, service, etc.

$$E_i = (t_j, a_k) \qquad (1)$$

Table 1. Event types

| Event types $(t_j)$ | | |
|---|---|---|
| Type 1 | Type 2 ... | Type m |

Table 2. Event attribute

| Event attributes ($a_k$) | | |
|---|---|---|
| Data types | Parameters | Values |

$E_i$ is the $i^{th}$ event, where $i \in \{1,2,3,...,n\}$, $t_j$ is the $j^{th}$ event type, $a_k$ is the $k^{th}$ event attribute, $da_x$ is the data type, $p_y$ is the parameter, $v_z$ is attribute value

$$t_j \in \{t_1, t_2, t_3, ...,t_m\} \quad (2)$$

$$a_k=(da_x,p_y,v_z) \quad (3)$$

$$da_x \in \{da_1, da_2, da_3, ...,da_k\} \quad (4)$$

$$p_y \in \{p_1, p_2,p_3, ...,p_l\} \quad (5)$$

$$v_z \in \{v_1, v_2 ,v_3, ...,v_x\} \quad (6)$$

$E_i$ occurs at a particular time is given by $E_i(t) = \begin{cases} 1; \text{Event has occured} \\ 0; \text{Otherwise} \end{cases}$

### 5.2. Function of condition module

Observe and receive an event, look for the rule that matches inputs and fire the rule as soon as condition hold. Condition module is 2-tuple consists of event details and condition attributes, $C_l=(d_x, c_m)$, Where $C_l$ is the $l^{th}$ condition, $d_x$ is the $x^{th}$ event details, $c_m$ is the $m^{th}$ condition attribute

$$d_x=(E_i,t_j,a_k) \quad \text{and} \quad c_m=(c_p,a_q,o_r, r_t) \quad (7)$$

Table 3. Event details

| Event details ($d_x$) |
|---|
| $E_i$, $t_j$, $a_k$ |

Table 4. Condition attributes

| Condition attributes($c_m$) | | | |
|---|---|---|---|
| Condition types | Arguments | Operators | Results |

### 5.3. Function of action module

- ➢ Based on event and condition an operation to be carried out.
- ➢ Action is 1-tuple consists of action attributes.
- ➢ Action $A_n=(at_p)$, where $A_n$ is the $n^{th}$ action, $at_p$ is the $p^{th}$ action attribute.
- ➢ When $\{E_i\}$ occurs if $\{C_l \text{ true}\}$ then $\{\text{Execute } A_n\}$.
- ➢ ECA scheme is distributed across the network i.e $R:\{E_i,C_l\} \rightarrow A_n$.

Table 5. Action attributes

| Action attributes ($at_p$) | | |
|---|---|---|
| $E_i$ | Condition result | Decision making |

**Algorithm1:** Algorithm for dynamic routing decision using ECA scheme

1: Begin
2: **Input:** Set of events
3: **Output:** Dynamic routing decision
4: **if**(Event has occurred in Ubiquitous network) **then**
5:     Split every event into event type and event attributes
6:     Apply logical condition
7:     Make runtime decision
8: **else**
9:  Do not make dynamic routing decision
10: **end if**
11: End

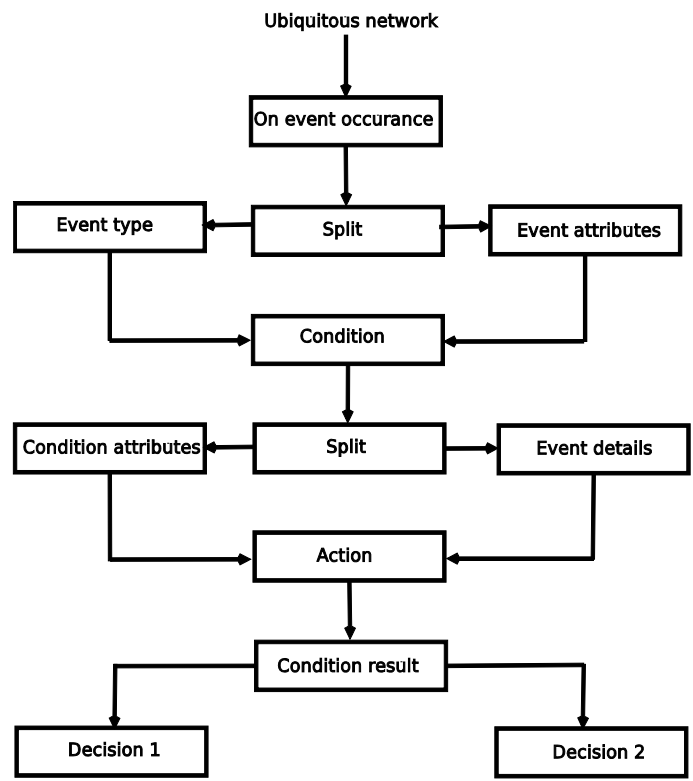

Figure 6. Event processing in ubiquitous network

## 5.4. ECA scheme based Routing in Ubiquitous Network

Table 6. Event(E1): Prepare route request and Event type(t1): Request

| Event attributes($a_k$) | | |
|---|---|---|
| Data types | Parameters | Values |
| GUID | Event unique ID | 9 |
| datetime | Date&time of event occurred | 10-03-2015 at 1:00pm |
| int | Packet type | RREQ=1 |
| bool | Join flag | Set J=1 or 0 |
| bool | Repair flag | Set R=1 or 0 |
| bool | Gratuitous flag | Set G=1 or 0 |
| bool | Destination only flag | Set D=1 or 0 |
| int | Hop count | Initial value=0 |
| int | Unique RREQ ID | 4 |
| string*ip_address | Destination IP address | 10.32.21.1 |
| int | Destination sequence # | 13 |
| string*ip_address | Source IP address | 10.32.21.83 |
| int | Source sequence number | 5 |

Table 7. Condition(C1): Event attributes($d_x$) are obtained from event module

| Condition attributes($c_m$) | | | |
|---|---|---|---|
| Condition types | Arguments | Operators | Results |
| Condition1: Checking valid path | Status | ?: | No valid path exist toward destination node |
| Condition2: Prepare RREQ message and setting sequence# | Set | += | Prepare RREQ message & set sequence number +=1 |

Table 8. Action(A1): Event and condition details are obtained

| Action attributes($at_p$) | | |
|---|---|---|
| Event | Condition result | Decision making |
| Prepare route request | Node does not have route to destination &&Prepare RREQ and set seq # += 1 | RREQ message is prepared and ready to broadcast |

**ECA Rule for Event(E1) : When**{Prepare route request event occurs} **If**{(Node does not have valid route) && (Sequence number+= 1)} **Then** {RREQ message is prepared and ready to broadcast}

Table 9. Event(E2):Generate route reply and Event type(t2): Request

| Event attributes($a_k$) | | |
|---|---|---|
| Data types | Parameters | Values |
| GUID | Event unique ID | 23 |
| datetime | Date&time of event occurred | 11-04-2015 at 2:00pm |
| int | Packet type | RREP=2 |
| bool | Repair flag | Set R=1 or 0 |
| bool | Acknowledgement flag | Set A=1 or 0 |
| bool | Prefix size | PS=00000 |
| int | Hop count | Initial value=0 |
| string*ip_address | Destination IP address | 10.32.21.83 |
| int | Destination sequence # | 2 |
| string*ip_address | Source IP address | 10.32.21.1 |
| time_t | Lifetime | 25msec |

Table 10. Condition(C2): Event attributes($d_x$) are obtained from event module

| Condition attributes($c_m$) | | | |
|---|---|---|---|
| Condition types | Arguments | Operators | Results |
| Condition1:Destination node | Status | != | Destination node that has active route |
| Condition2:Sequence number | Verify | > | Sequence number must be greater than that contained in RREQ message |

Table 11. Action(A2): Event and condition details are obtained

| Action attributes($at_p$) | | |
|---|---|---|
| Event | Condition result | Decision making |
| Generate route reply | Destination node that has active route && sequence # is greater than RREQ | Reverse route reply to source node |

**ECA Rule for Event(E2): When**{Generate route reply event occurs} **If**{(Destination node that has active route) && (Sequence# > RREQ message)} **Then** {Reverse route reply to source node}

Table 12. Event(E3):Route link has been broken and Event type(t3): Notification

| Event attributes($a_k$) | | |
|---|---|---|
| Data types | Parameters | Values |
| GUID | Event unique ID | 25 |
| datetime | Date&time of event occurred | 14-04-2015 at 5:00pm |
| int | Packet type | RERR=3 |
| bool | No delete flag | Set N=1 or 0 |
| int | Destination count | Initial value=0 |
| string*ip_address | Destination IP address | 10.32.21.51 |
| int | Destination sequence # | 6 |

Table 13. Condition(C3): Event attributes($d_x$) are obtained from event module

| Condition attributes($c_m$) | | | |
|---|---|---|---|
| Condition type | Argument | Operator | Result |
| Condition1:Link fail | Check | ?: | Invalidate the route |

Table 14. Action(A3): Event and condition details are obtained

| Action attributes($at_p$) | | |
|---|---|---|
| Event | Condition result | Decision making |
| Route link has been broken | Invalidate the route | List the affected nodes |

**ECA Rule for Event(E3): When**{Route link has been broken event occurs} **If**{Invalidate the route} **Then** { List the affected nodes}

Table 15. Event(E4):Generating route reply-ack and Event type(t4): Request

| Event attributes($a_k$) | | |
|---|---|---|
| Data types | Parameters | Values |
| GUID | Event unique ID | 14 |
| datetime | Date&time of event occurred | 21-04-2015 at 2:00pm |
| int | Packet type | RREP-ACK=4 |
| bool | Repair flag | Set R=1 or 0 |
| bool | Acknowledgement flag | Set A=1 or 0 |
| bool | Prefix size | PS=00000 |
| int | Hop count | Initial value=0 |
| string*ip_address | Destination IP address | 10.32.21.88 |
| int | Destination sequence # | 7 |
| string*ip_address | Source IP address | 10.32.21.13 |
| time_t | Lifetime | 25msec |

Table 16. Condition(C4): Event attributes($d_x$) are obtained from event module

| Condition attributes($c_m$) | | | |
|---|---|---|---|
| Condition type | Argument | Operator | Result |
| Condition1:Acknowledgement flag | Set | = | Use bidirectional link if A=1, |

Table 17. Action(A4): Event and condition details are obtained

| Action attributes($at_p$) | | |
|---|---|---|
| Event | Condition result | Decision making |
| Generating route reply-ack | Use bidirectional link if A=1, | Send route reply-ack |

**ECA Rule for Event(E4): When**{Generating route reply-ack} **If**{Bidirectional link flag A=1} **Then {** List the affected nodes}

### 5.5. ECA state transition diagram

- Event transition changes from one state to another state when initiated by a triggering event or condition.
- State machine is given by pentuple i.e. SM =($\sum$, S, $s_0$, $\delta$, F ).
- $\sum$ is the input which is considered as set of events $\sum$= {$E_i$}.
- S is the #of states i.e. state of a node S={1,2,...,n}.
- $s_0$ is the initial state, where an event has occurred.
- $\delta$ be the transition which is considered as condition.
- $\delta$ = {State$_{(old)}$ $\rightarrow$ State$_{(new)}$ ,Input$_{(condition)}$ $\rightarrow$ Output$_{(condition)}$ }.
- F is the final state, where an action will be taken.

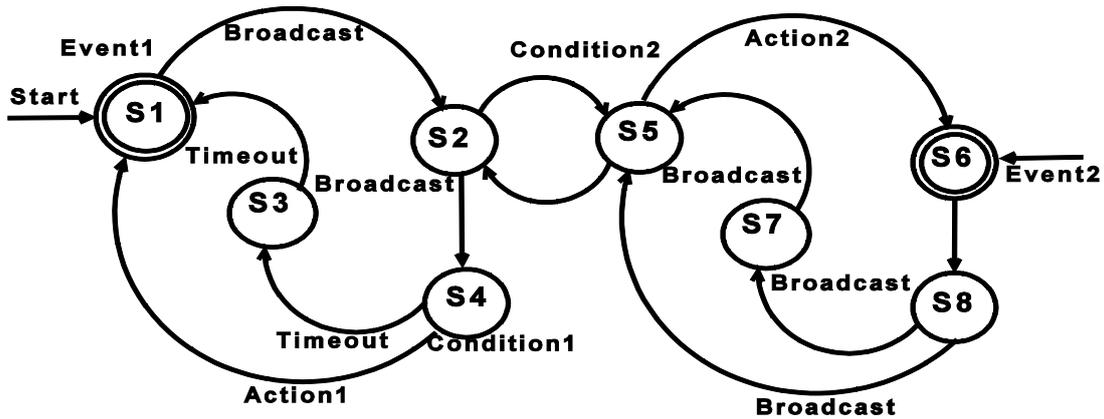

Figure 7. State transition diagram of an ECA Scheme

## 6. Case study: ECA scheme in Ubiquitous museum environment

Table 18. ECA scheme in ubiquitous museum environment

| Event-Condition-Action | | |
|---|---|---|
| Event | Condition | Action |
| User is looking for a route to visit exhibit | (Interest=Science) &&(Preference = Biology) | Provide route information about biology exhibit |
| Lunch time | (Preference=North-Indian) && (Time > 1 PM ) | Route to restaurant |
| High temperature in museum | Temperature $\geq 30^0$ C | Switch on AC |
| User blood pressure is low | BP < 90/60 | Provide shortest route to hospital |
| User is spending more time in front of exhibit | User's history says user is new to museum | Provide details information about exhibit |

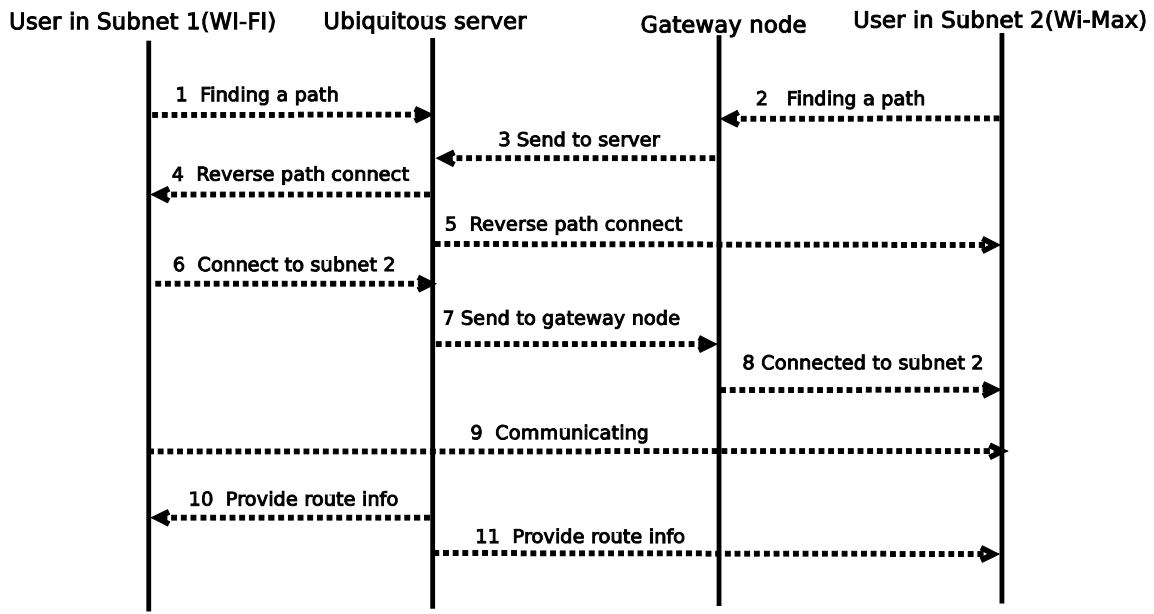

Figure 8. Sequence diagram to establish path between user and ubiquitous server

## 7. Simulation Environment

Proposed ECA scheme is implemented using C programming language and simulated in NS 2.34 simulator, by considering simulation parameters as depicted in table below. We have consider two different subnets such as Wi-Fi, MANETs in which random nodes are created and reverse path is established from ubiquitous server to the ubiquitous user, routing information is provided to the individual user based on interest, preferences.

Table 19. Simulation parameters

| Simulation parameters | |
|---|---|
| Parameters | Values |
| Nodes | 100 |
| Routing protocol | ECA-AODV |
| Transmission range | 30mts |
| Simulation time | 500s |
| Topology size | 25mX25m |
| Packet size | 512 Bytes |
| Mobility | Random |

## 8. Simulation Results

ECA scheme is simulated using NS 2.34 simulator and set of results are obtained as shown in the following figures. In fig #9, we have created 50 random nodes and simulated for 500ms for AODV and ECA-AODV protocol, ECA-AODV protocol achieve lower packet delivery latency than normal AODV protocol. Fig #10 in which 100 random nodes are created, different events are consider as an input for simulation by using 2 different routing protocols such as AODV and ECA-AODV respectively, after result execution, we conclude that data packet sending ratio of ECA-AODV is good in comparison with conventional AODV as and when event occur in the UbiNet.

In fig #11, we have shown mobility speed versus control byte transferred over data byte delivered with normal routing and ECA based routing. Fig #12 describes ECA scheme based RREQ and RREP message speed is high i.e. quicker response as and when event occur than the normal route request, route reply message and finally, fig #13 explain number of event processed per second by the node in ECA-AODV is better as and when number of node increases compared to the conventional AODV routing protocol

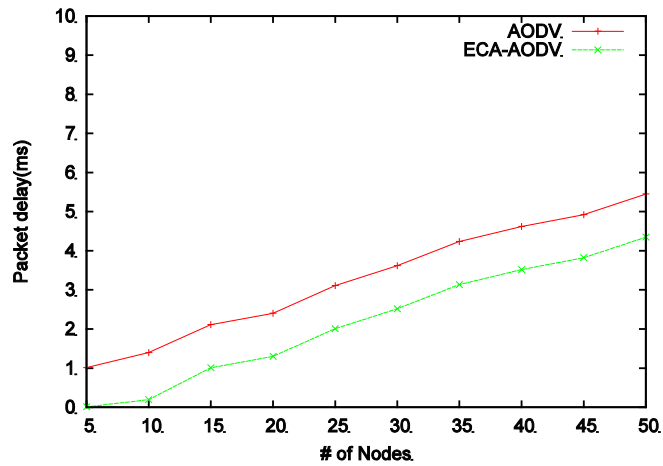

Figure 9. #of nodes in network Vs packet delivery

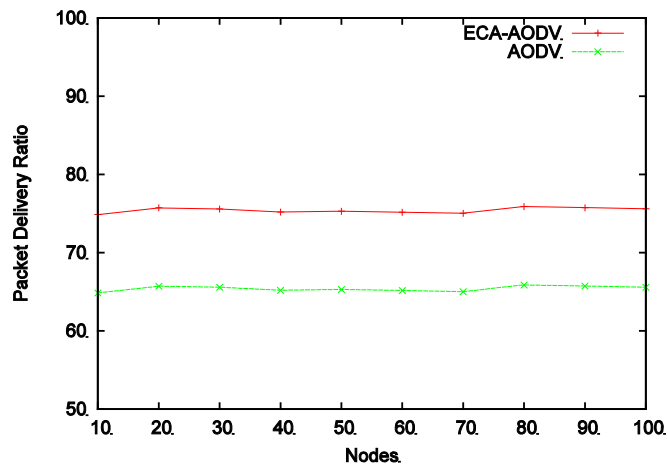

Figure 10. #of nodes Vs data packet delivery ratio

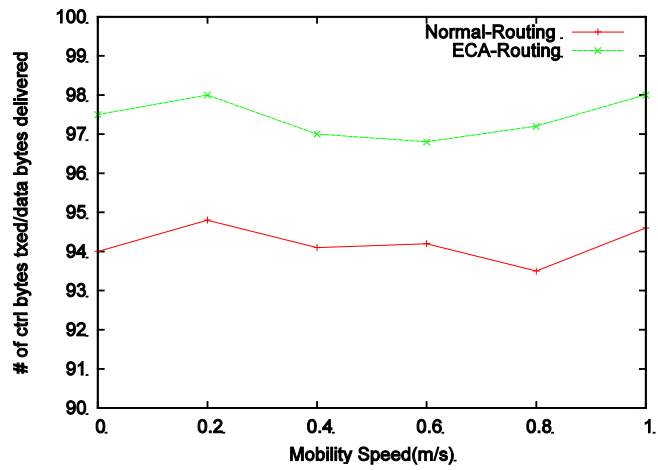

Figure 11. Mobility speed Vs #of ctrl bytes transferred/data bytes delivered

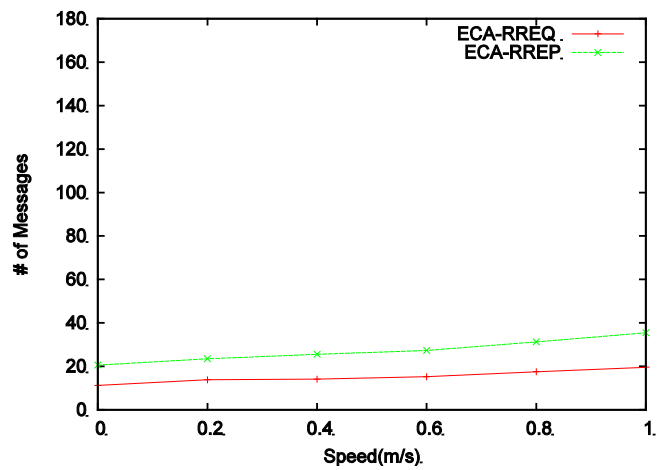

Figure 12. Speed Vs #of messages

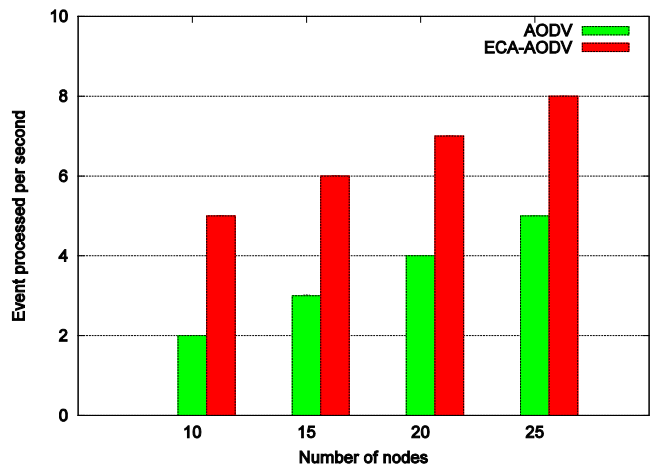

Figure 13. #of nodes Vs event processed per second

## 9. Conclusions

We have designed and simulated routing protocol using a novel ECA scheme in a UbiNet. Event module is designed using 2-tuples consists of event types and event attributes, condition module is designed using 2-tuples consists of event details and condition attributes and finally, an action module is designed using 1-tuple. We have considered ubiquitous museum environment as a case study to show the simulation of our proposed scheme, in which ubiquitous server provide uninterrupted connectivity and routing related information to the user as and when user move from one subnet to another subnet. However the proposed ECA scheme is flexible to supports dynamic network conditions in heterogeneous subnet and hence scheme is more effective as well as efficient as compared to non-ECA scheme in terms of parameters such as flexibility during runtime, easily adapt to the network dynamicity, quicker response as soon as event occur and easily adaptable to the types of network access technology used.

## Acknowledgments

We would like to thank our colleagues at Protocol Engineering and Technology unit, Department of ECE, Indian Institute of Science, Bangalore, India for their help and anonymous reviewers for their constructive and most valuable suggestions on improving the quality of the paper.

**Authors**

**Chandrashekhar Pomu Chavan** received his BE in Computer Science and Engineering from Guru Nanak Dev Engineering College, Bidar, Karnataka, India, and M.Tech Degree in Network and Internet Engineering from Sri Jayachamarajendra College of Engineering, Mysore, Karnataka, India, in 2005 and 2008 respectively. Currently he is pursuing his Ph.D degree on Ubiquitous Network under the guidance of Prof. Pallapa Venkataram in the Department of Electrical Communication Engineering at Indian Institute of Science, Bangalore, India. His research interests are in the areas of Ubiquitous 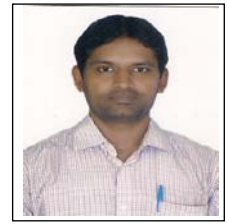 Computing, Pervasive Computing, Mobile Ad hoc Network, Context Aware System, and Routing Protocols.

**Prof. Pallapa Venkataram** received his Ph.D. Degree in Information Sciences from the University of Sheffield, England, in 1986. He is currently the chairman for center for continuing education, and also a Professor in the Department of Electrical Communication Engineering, Indian Institute of Science, Bangalore, India. Dr. Pallapa's research interests are in the areas of Wireless Ubiquitous Networks, Social Networks, Communication Protocols, Computation Intelligence applications in Communication Networks and Multimedia Systems. He is the holder of a Distinguished Visitor Diploma 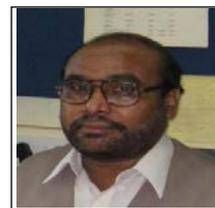 from the Orrego University, Trujillo, PERU. He has published over 200 papers in International/national Journals/conferences. Written three books: Mobile and wireless application security, Tata McGraw-Hill, Communication Protocol Engineering, Prentice-Hall of India (PHI), New Delhi, 2004 (Co-author: Sunil Manvi), and Multimedia: Concepts & Communication, Darling Kinderley(India) Pvt. Ltd., licensees of Pearson Education in South Asia, 2006. Edited two books: Wireless Communications for Next Millennium, McGraw-Hill, 1998, and Mobile Wireless Networks & Integrated Services, John Wiley & Sons(Asia) Pvt. Ltd., 2006(Co-editors: L.M.Patnaik & Sajal K. Das). Written chapters for two different books, and a guest editor to the IISc Journal for a special issue on Multimedia Wireless Networks. He has received best paper awards at GLOBECOM'93 and INM'95 and also CDIL (Communication Devices India Ltd) for a paper published in IETE Journal. He is a Fellow of IEE (England), Fellow of IETE (India) and a senior member of IEEE Computer Society.